\documentclass[12pt,preprint]{aastex}

\usepackage{natbib}
\usepackage{aas_macros}
\usepackage{color}
\usepackage{amsmath}
\usepackage{bm}

\newcommand{\myemail}{h.iijima@isee.nagoya-u.ac.jp}

\slugcomment{Accepted for publication in The Astrophysical Journal.}

\shorttitle{Twisted Chromospheric Jets}
\shortauthors{Iijima and Yokoyama}

\begin{document}


\title{
Three-dimensional Magnetohydrodynamic Simulation
of the Formation of Solar Chromospheric Jets
with Twisted Magnetic Field Lines
}


\author{H. Iijima}
\affil{
Institute for Space-Earth Environmental Research, Nagoya University,
Furocho, Chikusa-ku, Nagoya, Aichi 464-0814 Japan
}
\email{\myemail}
\and
\author{T. Yokoyama}
\affil{
Department of Earth and Planetary Science, The University of Tokyo,
7-3-1 Hongo, Bunkyo-ku, Tokyo 113-0033, Japan
}




\begin{abstract}
 This paper presents
 a three-dimensional simulation of chromospheric jets
 with twisted magnetic field lines.
 Detailed treatments of the photospheric radiative transfer
 and the equation of states allow us
 to model realistic thermal convection near the solar surface,
 which excites various MHD waves and
 produces chromospheric jets in the simulation.
 A tall chromospheric jet with a maximum height
 of 10--11 Mm and lifetime of 8--10 min is formed
 above a strong magnetic field concentration.
 The magnetic field lines are strongly entangled
 in the chromosphere, which helps
 the chromospheric jet to be driven by the Lorentz force.
 The jet exhibits oscillatory motion
 as a natural consequence of its generation mechanism.
 We also find that the produced chromospheric jet forms a cluster
 with a diameter of several Mm with finer strands.
 These results imply a close relationship
 between the simulated jet and solar spicules.
\end{abstract}





\section{Introduction}

Solar chromospheric jets are ubiquitously
observed in the lower layer of the solar atmosphere.
They involve various physical processes including
nonlinear amplification and mode conversion
of magnetohydrodynamic (MHD) shocks and waves,
as well as the thermodynamic and
inductive effects of partial ionization,
radiative energy exchange, and thermal conduction.

One of the most representative chromospheric jets is the solar spicule
\citep{1875leso.book.....S,1945ApJ...101..136R,1963ApJ...138..648B}.
This attractive phenomenon in the solar chromosphere
has been the subject of  various observational and theoretical studies
\cite[for a detailed review, see][]{1968SoPh....3..367B,
1972ARA&A..10...73B,2000SoPh..196...79S,2009SSRv..149..355Z,
2012SSRv..169..181T}.
Classical (or Type I) spicules are needle-like structures
observed at the solar limb,
with a maximum length of 4--10 Mm, lifetime of 1--7 min,
and maximum upward velocity of 20--100 km/s \citep{1972ARA&A..10...73B}.
Recently, the existence of a new class of spicules,
called Type II spicules, has been suggested.
Type II spicules exhibit a higher velocity of 50--150 km/s,
a shorter lifetime of 10--150 s \citep{2007PASJ...59S.655D},
and vigorous heating during the emergence
up to the transition region temperature
\citep{2014ApJ...792L..15P,2015ApJ...806..170S}.
Some authors have raised questions regarding
on the necessity of this new classification \citep{2012ApJ...750...16Z}
or the existence of the heating \citep{2016SoPh..291.2281B}.
In this study, we focus mainly on the properties of Type I spicules.

Various theoretical and numerical models have been suggested
to explain the generation mechanism
and vertical motion of solar spicules
\citep{2000SoPh..196...79S}.
These models are categorized by the energy source
required to lift the dense chromospheric plasma to the upper layer.
The acoustic wave model
\citep{1948ApJ...108..130T,1961PASJ...13..321U,
1961ApJ...134..347O,1964ApJ...140.1170P,
1982SoPh...75...99S,1982ApJ...257..345H}
is one such explanation.
Acoustic perturbation produced by the convective motion
\citep{2000ApJ...541..468S,2011ApJ...730L..24K,2016ApJ...827....7K}
is assumed to operate in the photosphere or chromosphere.
The upward acoustic wave is amplified owing to density stratification
\citep{1960PThPh..23..294O,1982SoPh...78..333S,2015ApJ...812L..30I}
and steepens into a shock wave.
When the shock wave reaches the transition region,
the transition region is elevated upward
by the shock-transition region interaction
\citep{1982ApJ...257..345H}.
In this model,
the vertically elongated chromospheric plasma
below the transition region
is observed as chromospheric jets.
Most theoretical models share the processes of
the acoustic/shock wave amplification
and shock-transition region interaction processes.
In the Alfv\'en wave model
\citep{1982SoPh...75...35H,1998A&A...338..729D,
1999ApJ...514..493K,2002A&A...393L..11J},
the nonlinear Alfv\'en wave is converted into the acoustic wave.
The shock-transition region interaction is slightly modified
by the existence of the magnetic field
in the magneto-acoustic shock wave \citep{1982SoPh...75...35H}.
Magnetic reconnection has also been suggested
as an energy source for acoustic wave generation
\citep{1969PASJ...21..128U,1969SvA....13..259P,
1971CoASP...3...33P,2009ApJ...702....1H,
2011PhPl...18k1210S,2017ApJ...836...24G}.
The magnetic reconnection model has the advantage
of chromospheric jets being accelerated
by both the shock wave and the Lorentz force
\citep{2013PASJ...65...62T}.

The horizontal oscillations of solar spicules
have also been reported in both imaging and spectroscopic observations
\citep{2009SSRv..149..355Z}.
The range of typical oscillation periods
reported in earlier studies is 1--7 min.
The typical amplitude ranges from 10 to 30 km/s.
This oscillation is assumed to be produced
by the transverse kink wave
\citep{2007Sci...318.1574D,2011ApJ...736L..24O}
or by the torsional Alfv\'en wave
\citep{2012ApJ...752L..12D}.
Both transverse \citep{1998ApJ...495..468S}
and torsional \citep{2012Natur.486..505W} waves
can be driven in the photosphere and chromosphere.
It is important to identify the wave mode and its propagation
to understand the energy transport into the upper atmosphere
\citep{2007Sci...318.1572E}.

Recent high-resolution observations have revealed
that an individual spicule has
multiple threads with widths of several hundreds of kilometers.
This value will be affected
by the spatial resolution of the instrument
\citep{2008ASPC..397...27S}.
They have also reported that the number of threads in a spicule
changes with time and interpreted this result
as a consequence of the spinning or torsional motion.
\cite{2010ApJ...714L...1S}
suggested that mini-filament eruptions
can explain this multi-threaded nature.
\cite{2014ApJ...795L..23S} discussed
the Kelvin-Helmholtz instability
caused by the transverse motion of a whole spicule
as the origin of the internal structure,
similar to the coronal loop simulations by \cite{2014ApJ...787L..22A}.

It is not an easy task to distinguish
the essential driving mechanism
from various and complex physical processes
in the solar chromosphere.
Although many theoretical models have been suggested,
we do not have a clear explanation of the origin of spicules.
The chromosphere is filled with shock waves
and the dynamic range of physical parameters is wide
owing to strong density stratification.
Various physical processes such as
the latent heat of partial ionization,
energy transport by radiation and thermal conduction,
and the collision between neutrals and ions
also contribute to chromospheric dynamics.
These various effects limit the identification
of the origin of chromospheric jets.

Realistic modeling of the solar chromosphere
by the radiation MHD simulations is
expected to overcome this difficulty.
Comparing radiation MHD simulations with observations,
\cite{2006ApJ...647L..73H} and \cite{2007ApJ...655..624D}
reported that the dynamic fibrils,
i.e., short chromospheric jets observed near active regions,
are driven by magneto-acoustic waves.
\cite{2009ApJ...701.1569M} suggested that
various mechanisms contribute to
the generation of small-scale jets.
\cite{2011ApJ...736....9M} reported that
one of the jets in their simulation
was similar to the Type II spicule
and was driven by the Lorentz force
with the flux emergence event.
\cite{2011ApJ...743..142H} carried out a detailed investigation
of acoustic wave propagation and jet formation.
These studies share the problem that
the tall ($>$ 6 Mm) chromospheric jets do not appear.
Recently, \cite{Martinez-Sykora1269} have suggested
that the ambipolar diffusion helps in increasing
the length of chromospheric jets driven by the tension force
and that the resulting jets can
reach a maximum height of $\approx8$ Mm.
\cite{2015ApJ...812L..30I} suggested
that lower coronal temperatures
can produce higher acoustically driven
chromospheric jets with a maximum height of 7 Mm.
However, the produced height is not enough
to explain the observed solar spicules
that sometimes exceed a maximum height of 10 Mm.

In this study,
we conduct a three-dimensional radiation MHD simulation
including the computational domain
from the upper convection zone to the lower corona.
The three-dimensional domain allows
plasma motion like the vertical vortex.
The purpose of this study is
to report the generation mechanism of these jets
and clarify the importance of the rotational motion
as a driver of chromospheric jets and solar spicules.

\section{Numerical Model}

The simulations are carried out using the numerical code
RAMENS\footnotemark[1] \citep{2015ApJ...812L..30I,2016PhDT.........5I}.
\footnotetext[1]{
RAdiation Magnetohydrodynamics Extensive Numerical Solver
}
The code solves the MHD equations
with gravity, Spitzer-type thermal conduction,
and radiative energy transport:
\begin{equation}
 \frac{\partial\rho}{\partial t}
  +\nabla\cdot\left(\rho\bm{V}\right)=0
\end{equation}
\begin{equation}
 \frac{\partial\left(\rho\bm{V}\right)}{\partial t}
 +\nabla\cdot\left[\rho\bm{V}\bm{V}
 +\left(P+\frac{B^2}{8\pi}\right)\underline{\mathrm{I}}
 -\frac{\bm{B}\bm{B}}{4\pi}\right]=\rho\bm{g}
\end{equation}
\begin{align}
 \begin{split}
  \frac{\partial e}{\partial t}
  &+\nabla\cdot\left[\left(e+P+\frac{B^2}{8\pi}\right)\bm{V}
  -\frac{1}{4\pi}\bm{B}\left(\bm{V}\cdot\bm{B}\right)\right]
  \\
  &=\rho\left(\bm{g}\cdot\bm{V}\right)
  +Q_\mathrm{cnd}+Q_\mathrm{rad}
 \end{split}
\end{align}
\begin{equation}
 \frac{\partial\bm{B}}{\partial t}
 +\nabla\cdot\left(\bm{V}\bm{B}
 -\bm{B}\bm{V}\right)=0
\end{equation}
Here, $\rho$ is the mass density,
$e=e_\mathrm{int}+\rho V^2/2+B^2/(8\pi)$
is the total energy density,
$e_\mathrm{int}$ is the internal energy density,
$\bm{V}$ is the velocity field,
$\bm{B}$ is the magnetic flux density,
$P$ is the gas pressure,
and $\bm{g}$ is the gravitational acceleration.
$Q_\mathrm{cnd}$ is the heating caused by the thermal conduction.
$Q_\mathrm{rad}$ is the combination of
optically thick radiative cooling,
computed in the gray approximation
in the photosphere and lower chromosphere,
and optically thin radiative cooling
in the upper chromosphere and corona.
The optically thin and thick cooling terms
are switched as a function of the column mass density.
The equation of states is computed under
the assumption of local thermodynamic equilibrium (LTE),
considering the six most abundant elements in the solar atmosphere.
The solar abundance is taken from \cite{2006CoAst.147...76A}.
The basic equations and numerical methods
are essentially the same as those in \cite{2015ApJ...812L..30I}.
Additional details of the numerical methods are
described in \cite{2016PhDT.........5I}.

We conduct a simulation with a three-dimensional numerical domain
spanning $9\times 9\times 16$ Mm$^3$,
including the upper convection zone with a depth of 2 Mm.
A uniform grid spacing of 41.7 km
in the horizontal ($X$ and $Y$) direction
and 29.6 km in the vertical ($Z$) direction is employed.
The horizontal boundary condition is periodic.
The top and bottom boundaries are open for flow.
The entropy of upward flow is fixed at the bottom boundary
so as to maintain thermal convection.
The thermal conductive flux from the top boundary
is imposed to maintain the coronal temperature to be higher than 1 MK.
To achieve a statistically evolved atmosphere
with limited computational resources,
we conduct the simulation run in multiple stages.
First, we impose a uniform vertical magnetic field of 10 G
on the sufficiently relaxed three-dimensional atmosphere
with a doubled horizontal grid spacing of 83.4 km.
We integrate this low-resolution simulation for three solar hours.
Next, we redefine the horizontal grid spacing as 41.7 km
and integrate the simulation for one solar hour.
We analyze the last 30 min of the simulation.

\section{Results}

\begin{figure}[!tp]
 \centering
 \plotone{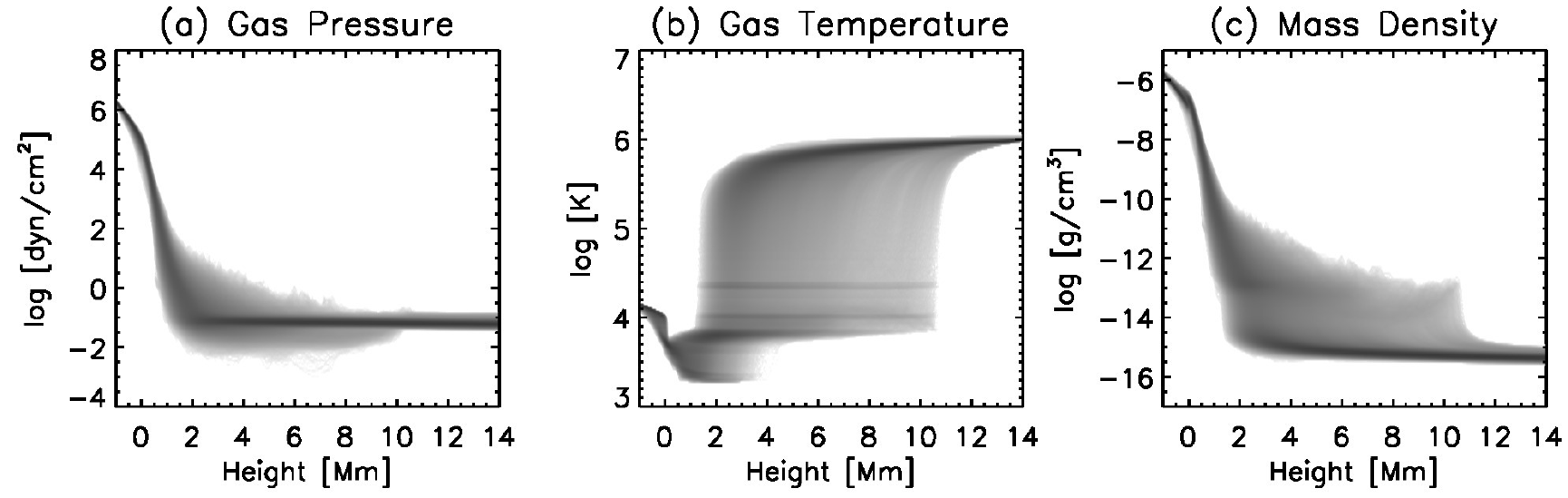}
 \caption{
 Vertical structure of the simulated atmosphere.
 The probability density functions (PDFs) of
 (a) gas pressure,
 (b) gas temperature,
 and (c) mass density are shown.
 }
 \label{fig:plhisth_m}
\end{figure}

The stratification of the simulated plasma
shows the vertically elongated chromosphere.
Figure \ref{fig:plhisth_m}
shows the vertical structure of the simulated atmosphere.
Because of the thermal conductive flux imposed at the top boundary,
the coronal temperature is maintained at approximately 1 MK.
The gas pressure and mass density are also uniform in the corona.
We also observe that the cool chromospheric plasma
with a mass density of $10^{-12}$--$10^{-14}$ g/cm$^3$
covers the height range of 2--10 Mm.
The result implies the existence of tall chromospheric jets
in the simulation as shown in the following sections.

\subsection{Morphology}

\begin{figure}[!tp]
 \centering
 \plotone{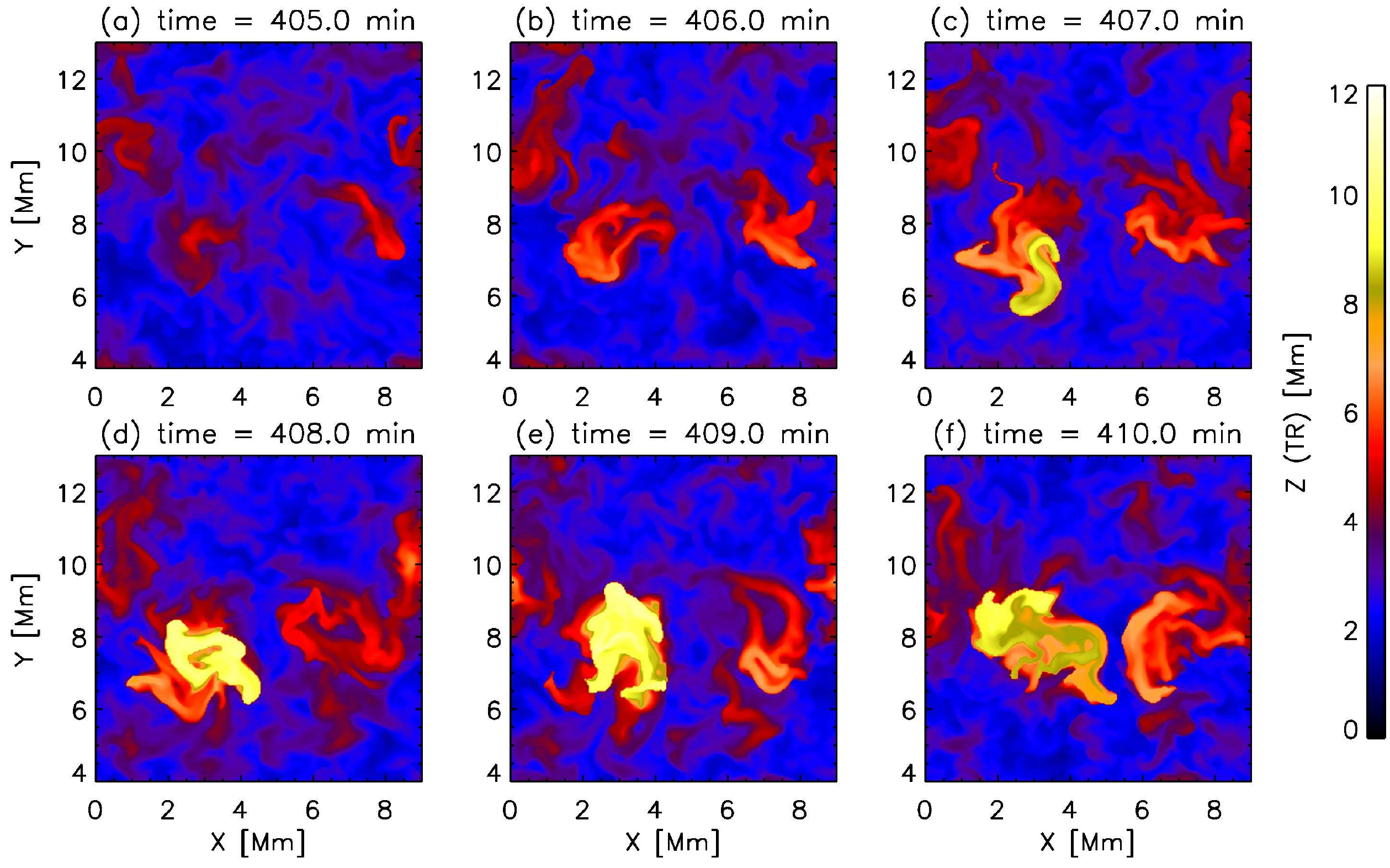}
 \caption{
 Height of the chromosphere-corona transition region,
 which is defined as the height at which $T=$ 40000 K.
 Jet-A is located at $(X, Y)\sim(3, 8)$ Mm.
 The variables of $Y>9$ Mm are calculated
 by assuming a periodic horizontal boundary condition.
 }
 \label{fig:plztr_m}
\end{figure}

The spatial structure of the tops of chromospheric jets
can be visualized by the geometrical height
of the transition region from the photospheric level.
Figure \ref{fig:plztr_m}
shows the time evolution of the height
of the chromosphere-corona transition region.
In this paper, we define the transition region as the height
at which the temperature is equal to 40000 K.
We find a tall jet that exceeds the maximum height of 10 Mm
at time $=409.0$ min near $(X, Y)\sim(3, 8)$ Mm.
Hereafter, we call this tall jet Jet-A.

We also identify the fine-scale horizontal structures
of chromospheric jets in Figure \ref{fig:plztr_m}.
Jet-A is a cluster of the fine-scale internal structures.
The internal structure gradually becomes complex
owing to the turbulent horizontal motion
caused by the counter-clockwise rotation of Jet-A.
This torsional motion is described in the next section.

The cluster of the fine-scale structures (Jet-A)
gradually widens horizontally during the evolution
(Figure \ref{fig:plztr_m}).
Jet-A initially has a horizontal diameter
of approximately 2 Mm at time $=407.0$ min.
The diameter reaches almost 4 Mm at time $=410.0$ min.
The spatial deviation among the internal fine-scale structures
increases during the evolution of the jet.

In this study, we focus on the properties and formation
of Jet-A as a representative chromospheric jet
in our simulation.
Jet-A is the tallest chromospheric jet
within the 30 min duration of the simulation.
We find several jets exceeding the height of 7 Mm
during the whole simulation run,
with similar rotating motions and fine-scale structures.

\begin{figure}[!tp]
 \centering
 \plotone{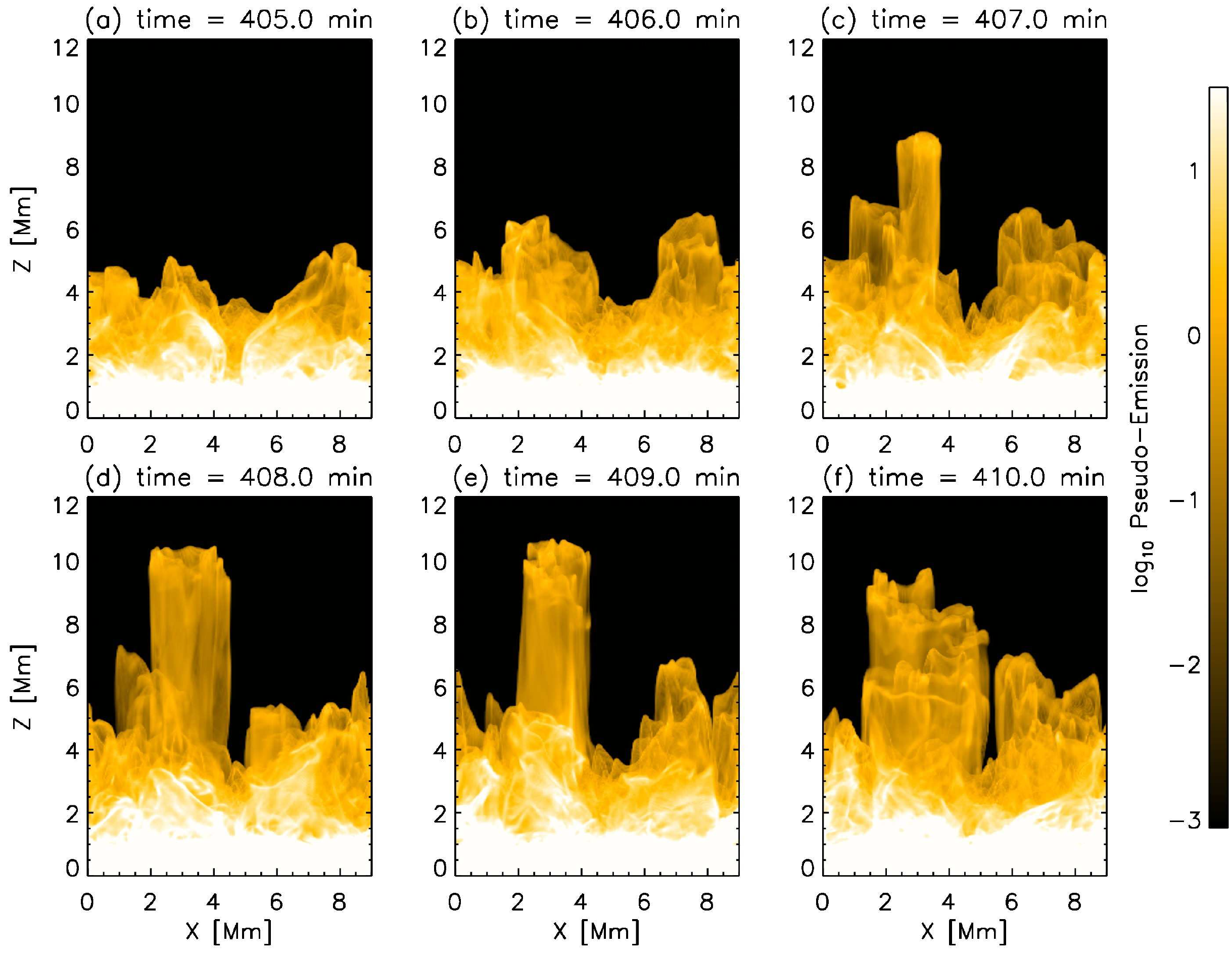}
 \caption{
 Six snapshots of the pseudo-emission
 defined in Eq. (\ref{eq:aemit})
 and observed from the negative $Y$-direction.
 Jet-A is located at $X\sim 3$ Mm.
 }
 \label{fig:plamity_m}
\end{figure}

For the visualization mimicking the observation at the solar limb,
we use pseudo-emission $\epsilon$
in the optically thin approximation
defined as
\begin{equation}
 \epsilon=\int n_\mathrm{e}n_\mathrm{H}G(T)\mathrm{d}l,
  \label{eq:aemit}
\end{equation}
where
$n_\mathrm{e}$ is the number density of electrons
and $n_\mathrm{H}$ is the number density of hydrogen nuclei.
The line integral is taken
along the line-of-sight of the pseudo-observation.
We assume a Gaussian form
of the contribution function $G(T)$, given as
\begin{equation}
 G(T)=C_0\exp
  \left[
   -\left(
     \frac{\log_{10} (T/T_c)}{\varDelta \log_{10} T}
    \right)^2
  \right],
  \label{eq:cntrb}
\end{equation}
where we set $\log_{10} T_c\ \mathrm{[K]}=4.2$ and
$\varDelta \log_{10} T=0.15$.
The normalization constant $C_0=10^{-28}$ cm$^{5}$ is taken
such that the pseudo-emission in Eq. (\ref{eq:aemit})
becomes unity (dimensionless)
for a plasma with a number density of
$n_\mathrm{e}=n_\mathrm{H}=10^{10}$ cm$^{-3}$
and a line-of-sight thickness of 1 Mm.
This contribution function is
sensitive to the temperature range of the upper chromosphere.

The pseudo-emission from the negative $Y$-direction
(Figure \ref{fig:plamity_m})
shows the possible appearance of Jet-A
when we observed at the solar limb.
The line integral in Eq. (\ref{eq:aemit})
is taken over the whole $Y$-direction for each $(X,Z)$.
Jet-A is located at $X\sim 3$ Mm.
The top of the jet reaches
a maximum height of approximately 10 Mm at time $=409.0$ min
(Figure \ref{fig:plamity_m}(e)).
We also find finer jetting strands
with a horizontal size of several hundreds of kilometers in Jet-A.
The maximum height of $\approx10$ Mm
and the existence of the fine-scale structure
are consistent with the findings in Figure \ref{fig:plztr_m}.

\begin{figure}[!tp]
 \centering
 \plotone{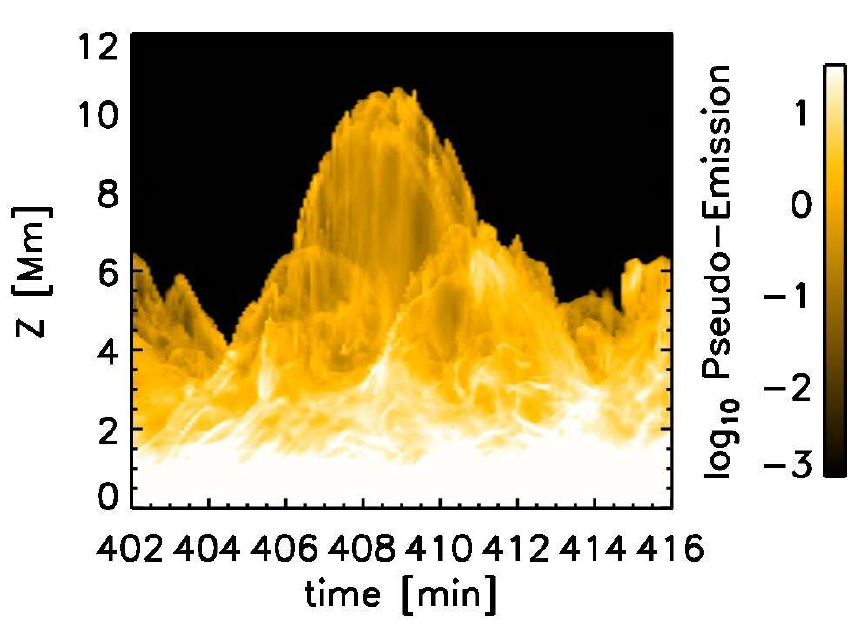}
 \caption{
 Time evolution of the pseudo-emission
 at the slit of $X=2.7$ Mm in Figure \ref{fig:plamity_m}.
 }
 \label{fig:plamity_tz}
\end{figure}

We put a vertical slit at $X=2.7$ Mm in Figure \ref{fig:plamity_m}
for a more quantitative description of the vertical motion of Jet-A.
Figure \ref{fig:plamity_tz}
shows the time evolution of the pseudo-emission of Jet-A.
The top of Jet-A exhibits a nearly parabolic trajectory.
The jet has a maximum elevation height of $\sim 7$ Mm
and a lifetime of $\sim 9$ min.
We estimate the maximum upward velocity of $\sim 50$ km/s
and deceleration of $\sim 200$ m/s$^2$
by assuming a purely parabolic trajectory.

\begin{figure}[!tp]
 \centering
 \epsscale{0.7}
 \plotone{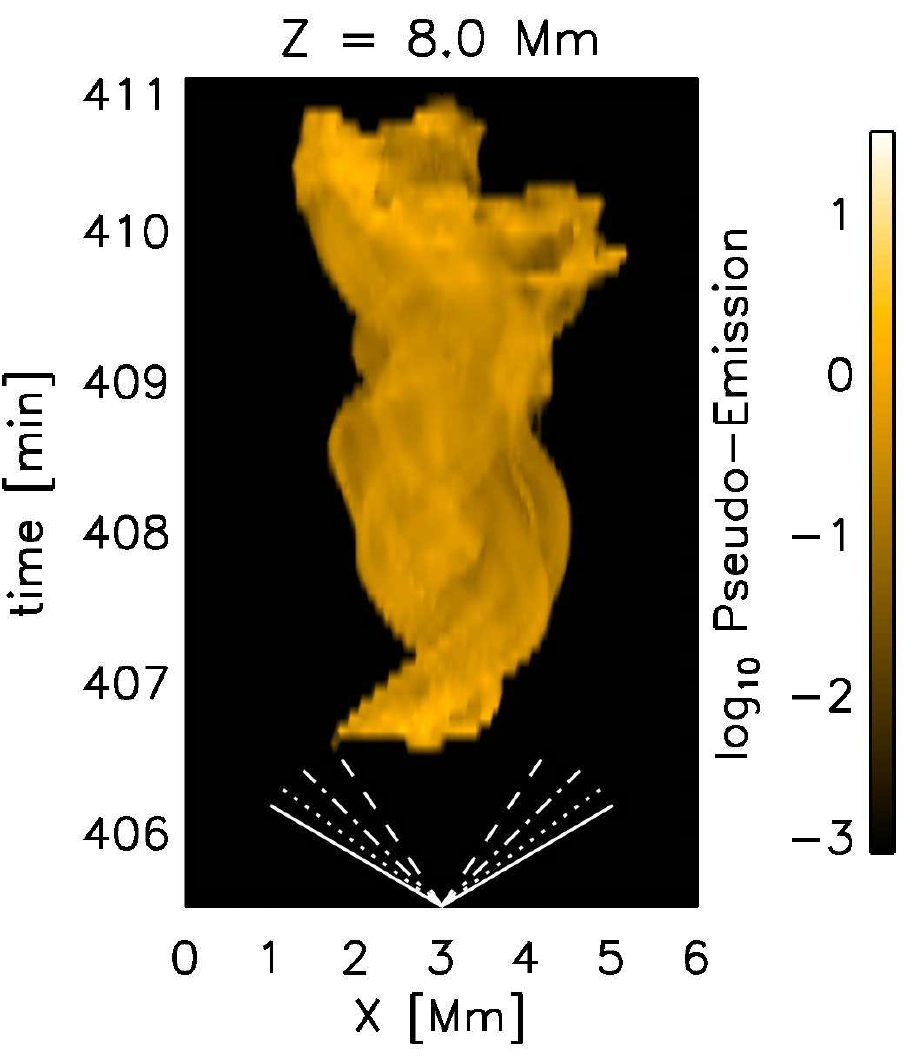}
 \caption{
 Time evolution of the horizontal slices of the pseudo-emission
 at the slit of $Z=8$ Mm in Figure \ref{fig:plamity_m}.
 The white lines in each plot indicate
 the constant-velocities
 of 20 km/s (dashed), 30 km/s (dash-dotted), 40 km/s (dotted),
 and 50 km/s (solid) along the $X$-direction.
 }
 \epsscale{1.0}
 \label{fig:plamity_xt}
\end{figure}

The pseudo-emission (Figure \ref{fig:plamity_m}) also clarifies
the apparent horizontal motion of the fine-scale strands in Jet-A.
We put a horizontal slit at $Z=8$ Mm
for a detailed description of the horizontal motion
(Figure \ref{fig:plamity_xt}).
We find wave-like patterns
with apparent velocities of 30--40 km/s,
displacements of $\approx1$ Mm,
and periods of 2--3 min.
These patterns correspond to
the horizontal displacement of fine-scale strands
that follows the rotating motion of Jet-A
which can be found in Figure \ref{fig:plztr_m}.

When we interpret Jet-A as a single jet
neglecting the internal structure,
the apparent horizontal size of Jet-A
changes in time within a period of several minutes
(Figure \ref{fig:plamity_m}).
This temporal change in horizontal width
is caused by the combination of
the rotating motion of Jet-A and
the temporal change of its internal horizontal structure,
as shown in Figure \ref{fig:plztr_m}.
The apparent horizontal size of Jet-A
gradually increases with the oscillation,
probably owing to the centrifugal force of the rotation,
as discussed in Section \ref{sec:discuss}.
Jet-A as a single jet also exhibits an apparent swaying motion.
The horizontal velocity ($\approx20$ km/s)
and the displacement of the center ($\approx0.5$ Mm)
of this large-scale swaying pattern 
are slightly smaller than those of each strand
because of the average effect.

\subsection{Structure of Velocity and Magnetic Field}

\begin{figure}[!tp]
 \centering
 \plotone{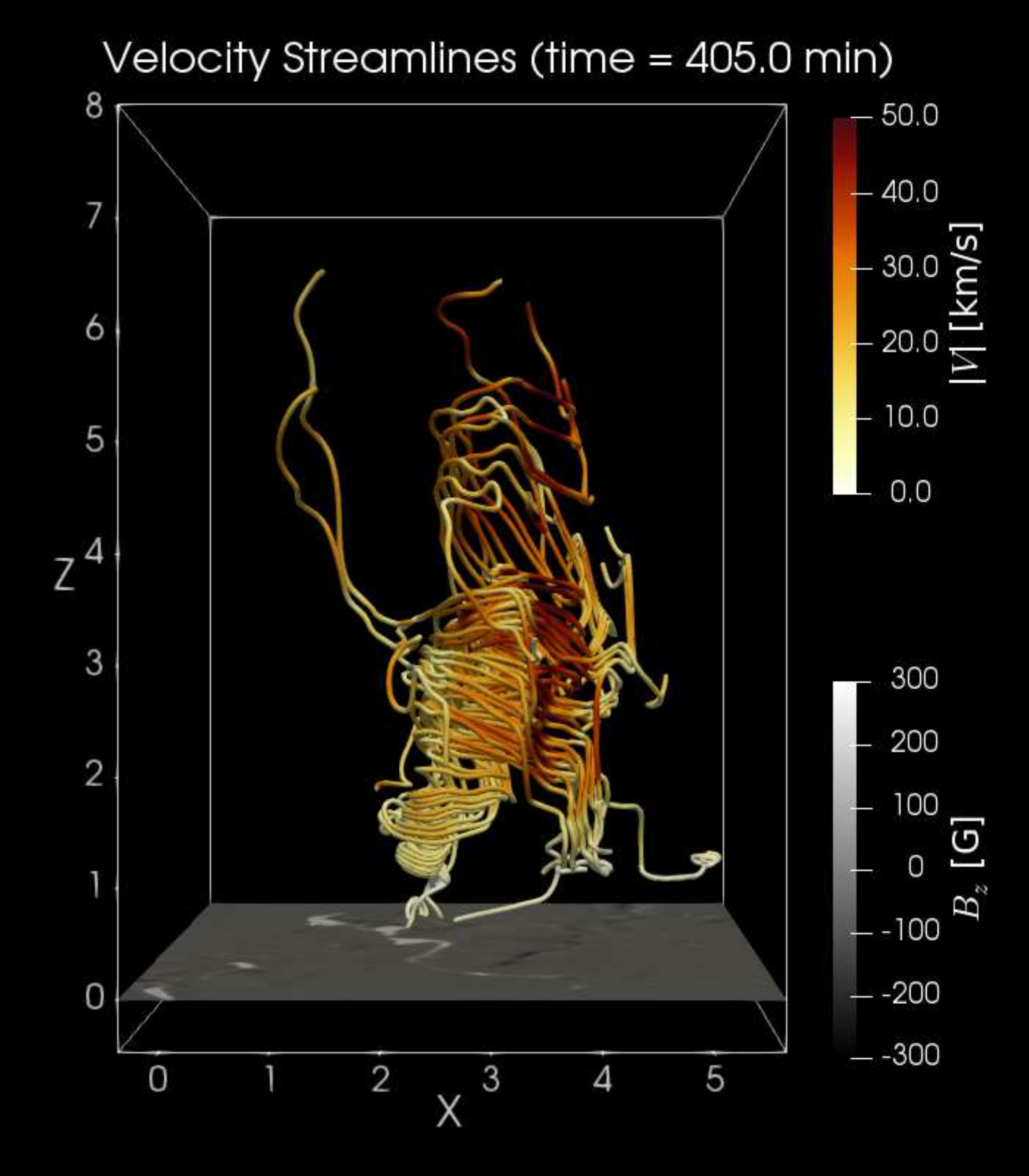}
 \caption{
 Three-dimensional structure of the velocity streamlines
 in a small box of $6\times6\times8.5$ Mm
 (including 0.5 Mm below $Z=0$ Mm)
 near the foot point of Jet-A located at $(X,Y)\sim(3,8)$ Mm.
 A snapshot at time of 405 min is shown.
 The horizontal slice of the vertical magnetic field
 at $Z=0$ Mm is also shown in gray scale.
 The color along the streamlines represents
 the magnitude of the velocity.
 (An animation of this figure is also available.)
 }
 \label{fig:pv12_0108}
\end{figure}

\begin{figure}[!tp]
 \centering
 \plotone{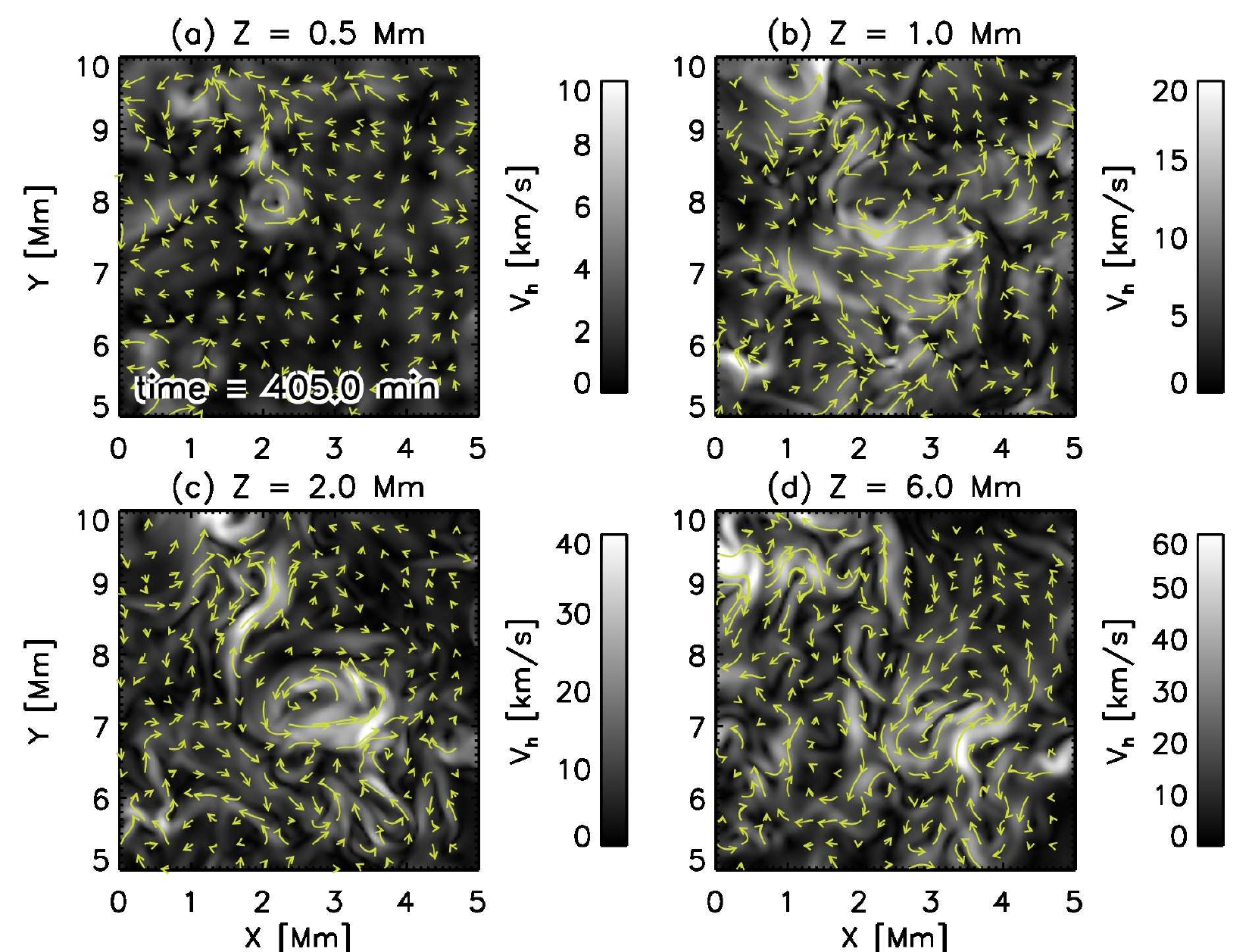}
 \caption{
 Horizontal slices of
 the absolute of the horizontal velocity
 at time $=$ 405.0 min.
 The horizontal slices are located at
 0.5 Mm (panel (a)), 1.0 Mm (panel (b)),
 2.0 Mm (panel (c)), and 6.0 Mm (panel (d)).
 The arrows in the right panels indicate the horizontal velocity field.
 The variables of $Y>9$ Mm are calculated
 by assuming a periodic horizontal boundary condition.
 }
 \label{fig:plxyt_z_m_vh}
\end{figure}

Near the root of Jet-A,
an anticlockwise vertical vortex is found in the chromosphere
(Figures \ref{fig:pv12_0108} and \ref{fig:plxyt_z_m_vh}).
This strong vortex extends from the photosphere to the corona.
The vortex is located at $(X,Y)\sim(2,8)$
near the temperature minimum (Figures \ref{fig:plxyt_z_m_vh}(a)).
The center of the vortex moves 
toward the positive $X$-direction and negative $Y$-direction
in the higher layers (Figures \ref{fig:plxyt_z_m_vh}(b)--(d)).
The result indicates the inclination
of the vertical vortex from the vertical axis.
The strength of this chromospheric vortex
is about 10--40 km/s depending on the measured height
(Figures \ref{fig:plxyt_z_m_vh}(b) and (c)).
The vortex becomes wider in the higher region.

\begin{figure}[!tp]
 \centering
 \plotone{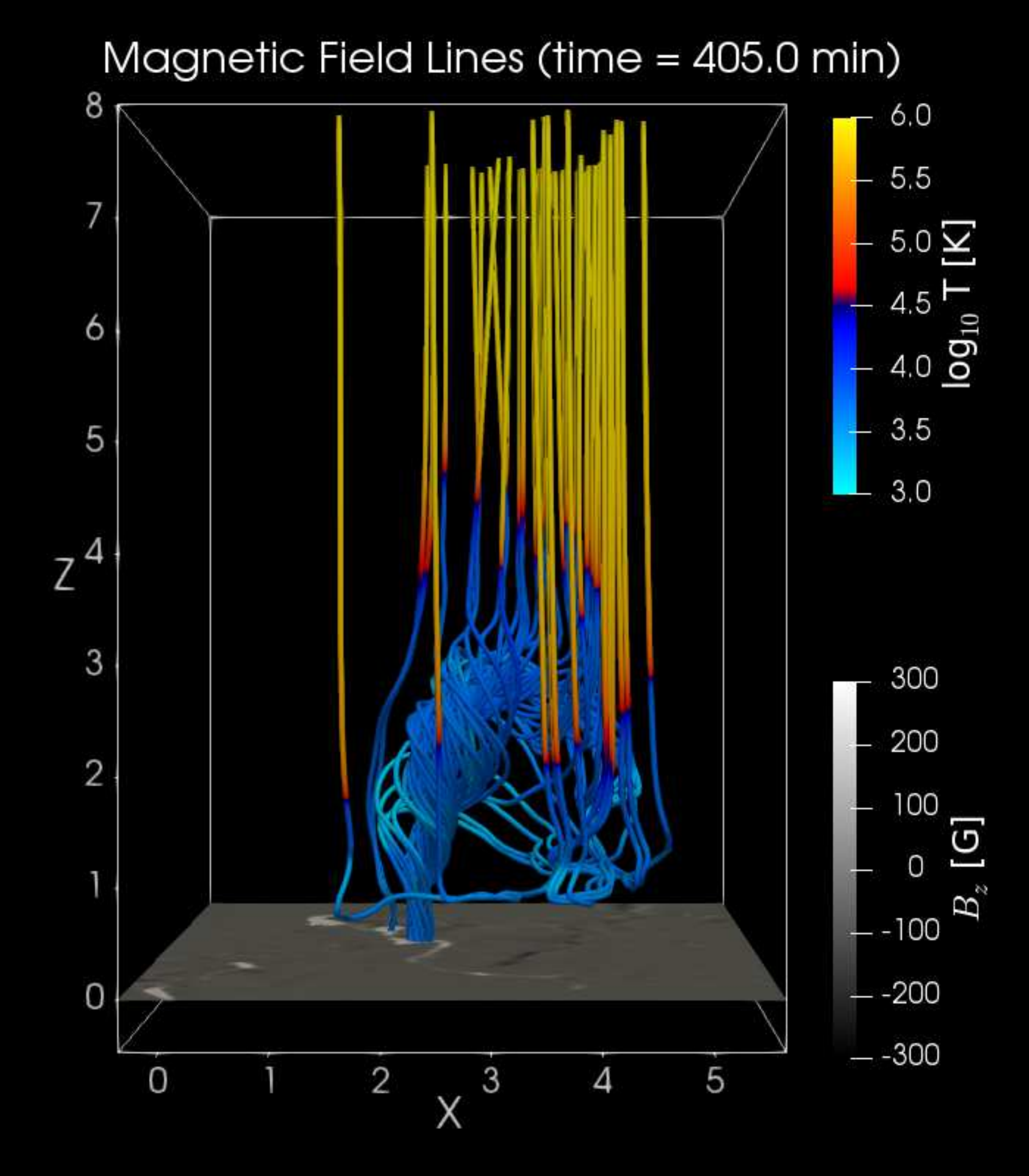}
 \caption{
 Same as Figure \ref{fig:pv12_0108},
 but showing the three-dimensional structure
 of the magnetic field lines.
 The color along the magnetic field lines indicates the gas temperature.
 (An animation of this figure is also available.)
 }
 \label{fig:pv11_0108}
\end{figure}

The magnetic field lines near the foot point of Jet-A
are highly twisted in the chromosphere (Figure \ref{fig:pv11_0108}).
We also find that the magnetic field lines
are deformed near $(X,Z)\sim(4,1)$.
The axis of the twist is inclined
from the vertical axis
as observed in the vortex structure
(Figures \ref{fig:pv12_0108} and \ref{fig:plxyt_z_m_vh}).
This result is interesting because the previous reports
of the twisted magnetic field lines are relatively rare,
although the torsional motion in the chromosphere
has sometimes been reported in the simulations and observations
\citep{2012Natur.486..505W,2013ApJ...770...37K,2013ApJ...776L...4S}.
A more detailed discussion on the comparison
with the previous studies is given in Section \ref{sec:discuss}.

\begin{figure}[!tp]
 \centering
 \plotone{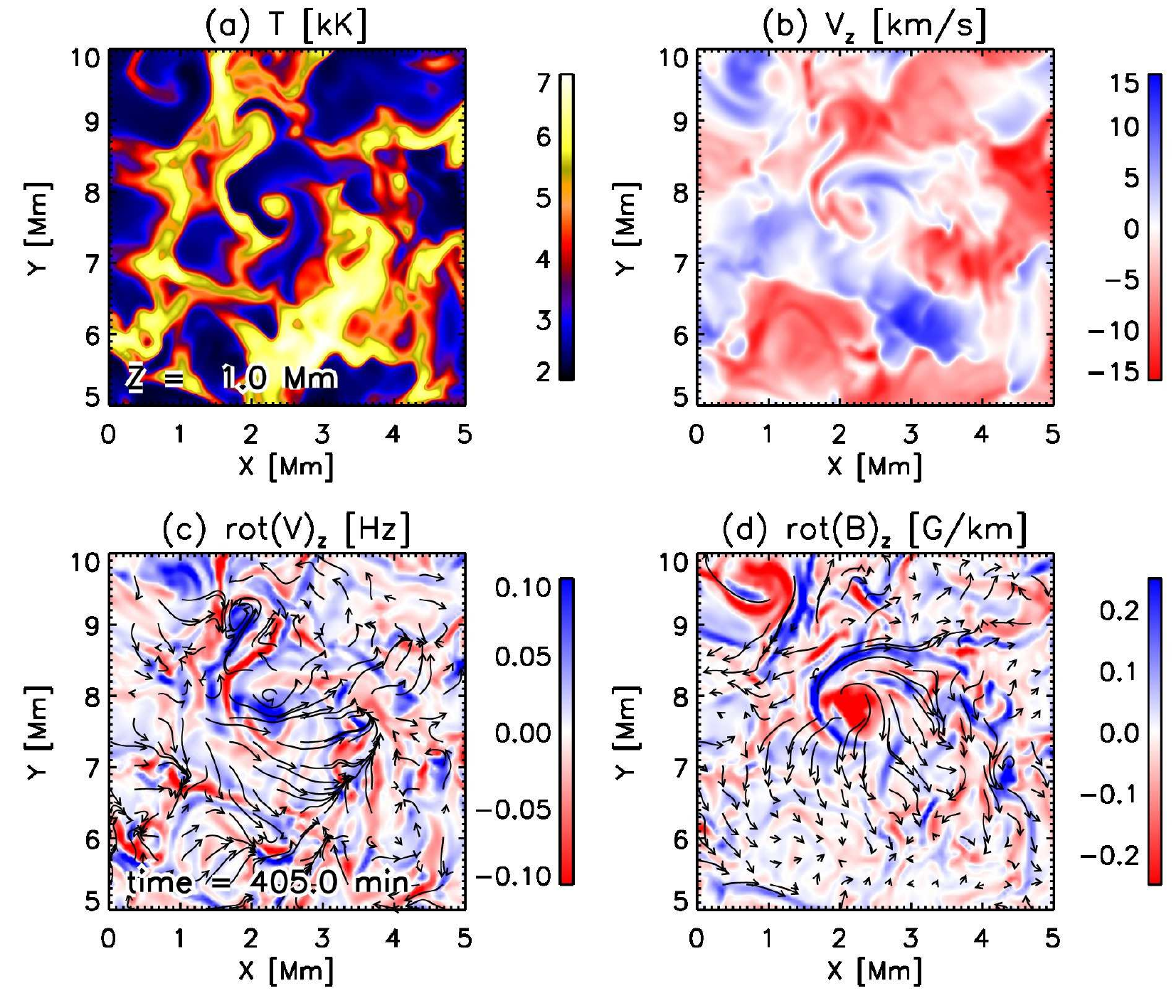}
 \caption{
 (a) Temperature,
 (b) vertical component of the velocity field,
 (c) vertical component of the vorticity,
 and (d) vertical component of the rotation of the magnetic field
 in the horizontal plane at $Z=1$ Mm
 and time $=$ 405.0 min.
 The arrows in panels (c) and (d) indicate
 the horizontal velocity 
 and magnetic fields, respectively.
 The variables of $Y>9$ Mm are calculated by
 assuming a periodic horizontal boundary condition.
 }
 \label{fig:plxyt_z_m_0105_zoom.0108}
\end{figure}

The twist of the magnetic field lines in the chromosphere
is caused by the vortex motion.
Figure \ref{fig:plxyt_z_m_0105_zoom.0108} shows
the horizontal slice at $Z=1$ Mm at the foot point of Jet-A.
The small-scale swirl is found at $(X,Y)\sim(2,8)$ Mm.
The lifetime of this swirl event is longer than 5 min.
The swirl has a hot center with a downward flow
and a cool edge with an upward flow.
The vertical vorticity has a sign opposite
to that of the vertical component
of the curl of the magnetic field vector
(Figures \ref{fig:plxyt_z_m_0105_zoom.0108}(c) and (d)),
which is consistent with
the upward-propagating torsional Alfv\'en wave.
This result indicates that this swirl motion causes
the twisted structure of the magnetic field lines
in Figure \ref{fig:pv11_0108}.

\begin{figure}[!tp]
 \centering
 \plotone{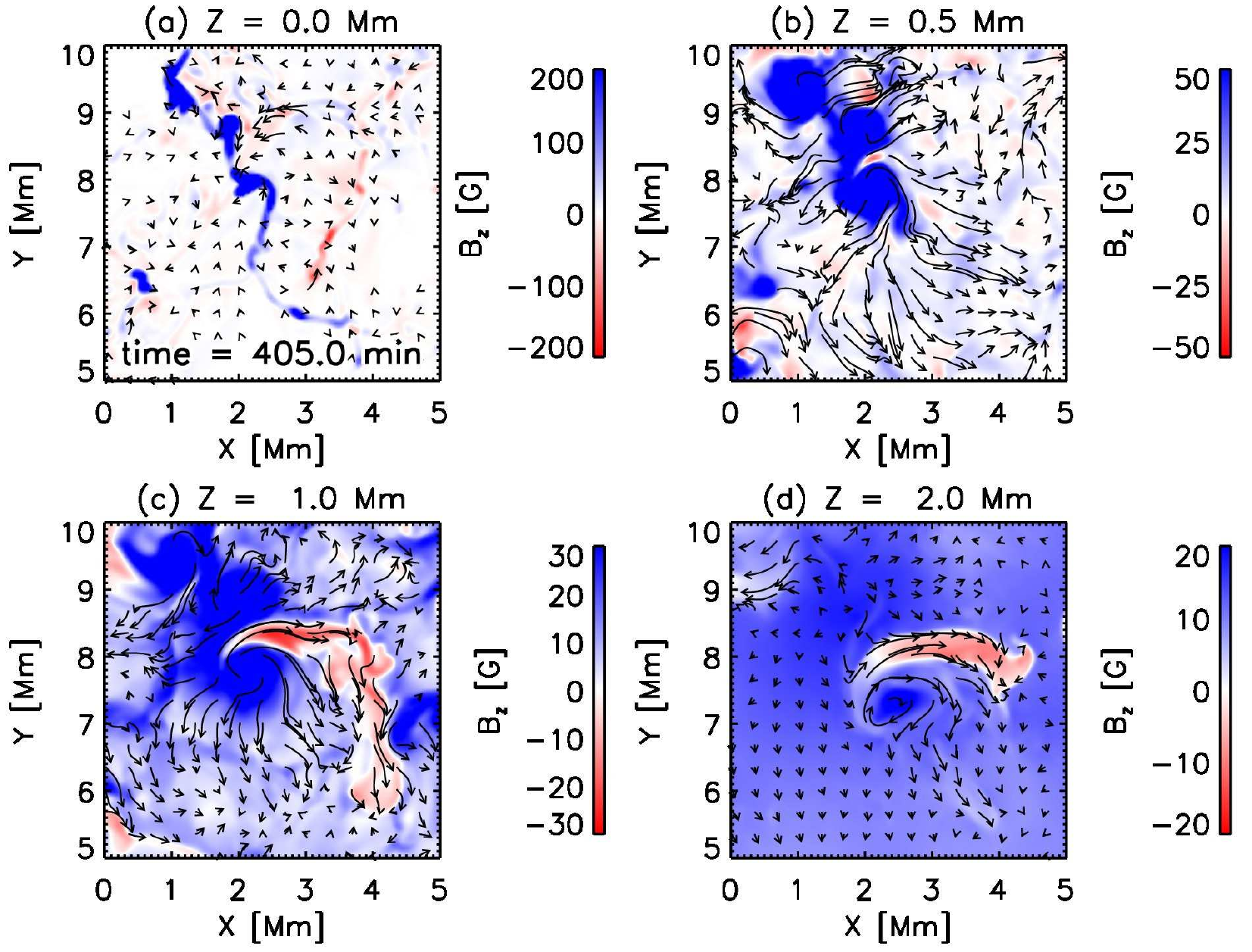}
 \caption{
 Vertical magnetic field
 at the four different layers of
 $Z=$ (a) 0.0, (b) 0.5, (c) 1.0, and (d) 2.0 Mm
 at time $=$ 405.0 min.
 Arrows indicate the horizontal component of the magnetic field.
 The variables of $Y>9$ Mm are calculated by
 assuming a periodic horizontal boundary condition.
 }
 \label{fig:plxyt_z_m_bz.0108}
\end{figure}

We find a negative patch of the vertical magnetic field
near the foot point of Jet-A
in the middle of the chromosphere.
Figure \ref{fig:plxyt_z_m_bz.0108} shows
the vertical magnetic field at various heights.
The magnetic field lines through Jet-A
originate from the photospheric magnetic flux concentration
located at $(X,Y)\sim(2,8)$ Mm in Figure \ref{fig:plxyt_z_m_bz.0108}(a).
We find a vertical magnetic field with the opposite polarity,
which is located with a finite offset against
the axis of the swirling motion
in Figure \ref{fig:plxyt_z_m_0105_zoom.0108}.
This opposite-polarity patch in the chromosphere
(at $Z=$ 1.0 and 2.0 Mm; Figures \ref{fig:plxyt_z_m_bz.0108}(c)
and \ref{fig:plxyt_z_m_bz.0108}(d), respectively)
exists neither near the temperature minimum
(at $Z=$ 0.5 Mm; Figure \ref{fig:plxyt_z_m_bz.0108}(b))
nor at the photospheric surface
(at $Z=$ 0.0 Mm; Figure \ref{fig:plxyt_z_m_bz.0108}(a)).
The direction of the horizontal magnetic field
in this negative magnetic patch is the same
as the nearby positive patch
(Figures \ref{fig:plxyt_z_m_bz.0108}(c)
and \ref{fig:plxyt_z_m_bz.0108}(d)).

\begin{figure}[!tp]
 \centering
 \plotone{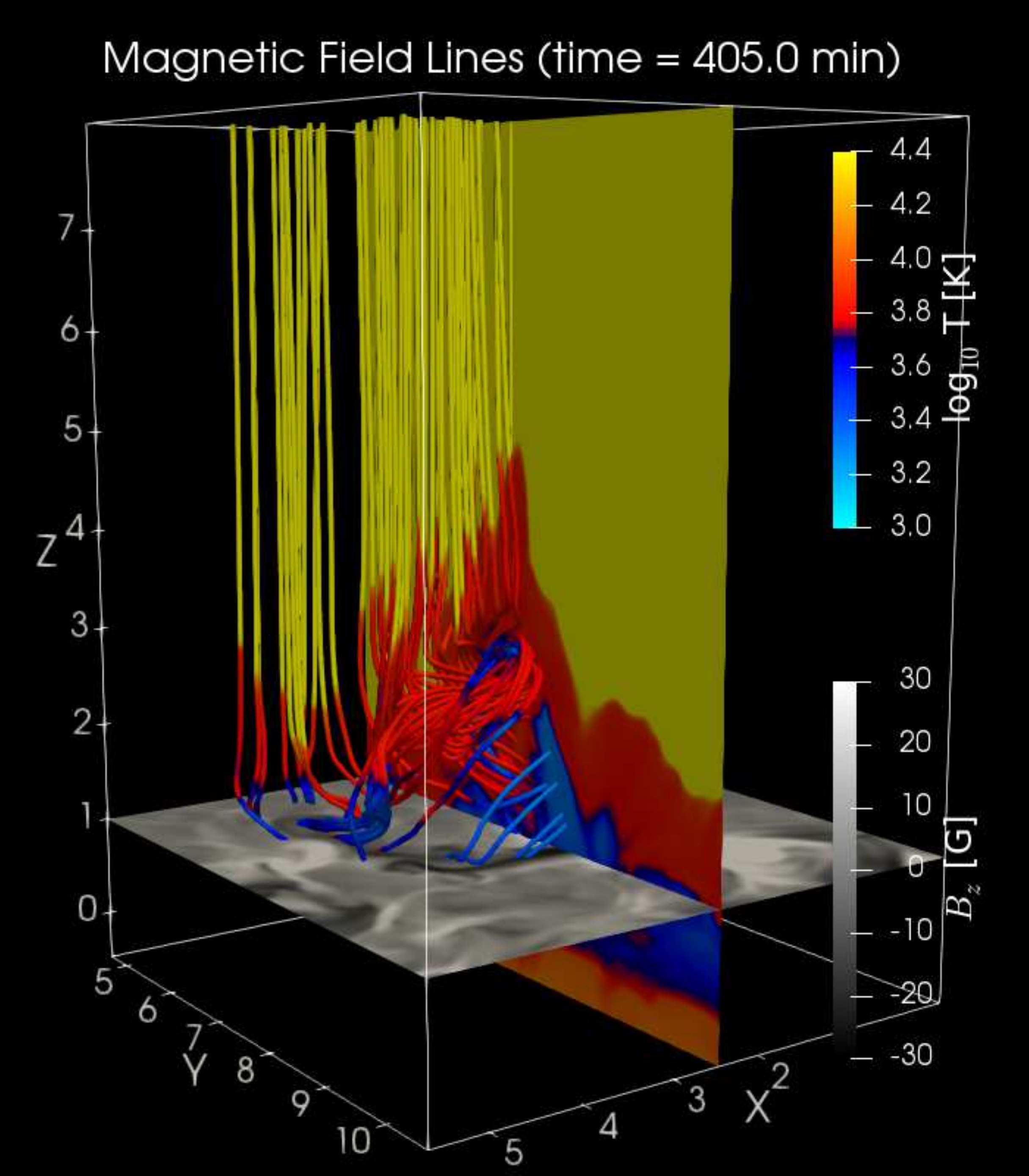}
 \caption{
 Three-dimensional structure of the magnetic field lines.
 The horizontal slice of the vertical magnetic field
 at $Z=1$ Mm is shown in gray scale.
 Gas temperature on the $(Y,Z)$-plane at $X=2.5$ Mm is also shown.
 (An animation of this figure is also available.)
 }
 \label{fig:pv13_0108}
\end{figure}

The negative magnetic patch in the chromosphere
(Figure \ref{fig:plxyt_z_m_bz.0108}(c))
is a deformed part of the field lines
connected to the positive patch in the photosphere.
Figure \ref{fig:pv13_0108} shows
the three-dimensional topology of the magnetic field lines
with the superposed horizontal slice at $Z=1$ Mm.
The negative-polarity magnetic patch
found in Figures \ref{fig:plxyt_z_m_bz.0108}(c) and (d)
is a part of the positive magnetic patch
located at $(X,Y,Z)\sim(2,8,0)$ Mm
in Figure \ref{fig:plxyt_z_m_bz.0108}(a).

The torsional rotation near the foot point of Jet-A,
which inclines slightly from the vertical axis,
helps in producing these deformed magnetic field lines.
The swirling magnetic field lines
drag the heavy and cool plasma upward in the chromosphere
(Figure \ref{fig:pv13_0108}).
This dragged heavy plasma deforms the magnetic field lines
and produces a negative magnetic patch
in the middle chromosphere.
These deformed magnetic field lines
and the corresponding dragging motion
also contribute to driving Jet-A,
as will be discussed in Section \ref{sec:driver}.

\subsection{Driving Mechanism of Chromospheric Jets}\label{sec:driver}

\begin{figure}[!tp]
 \centering
 \plotone{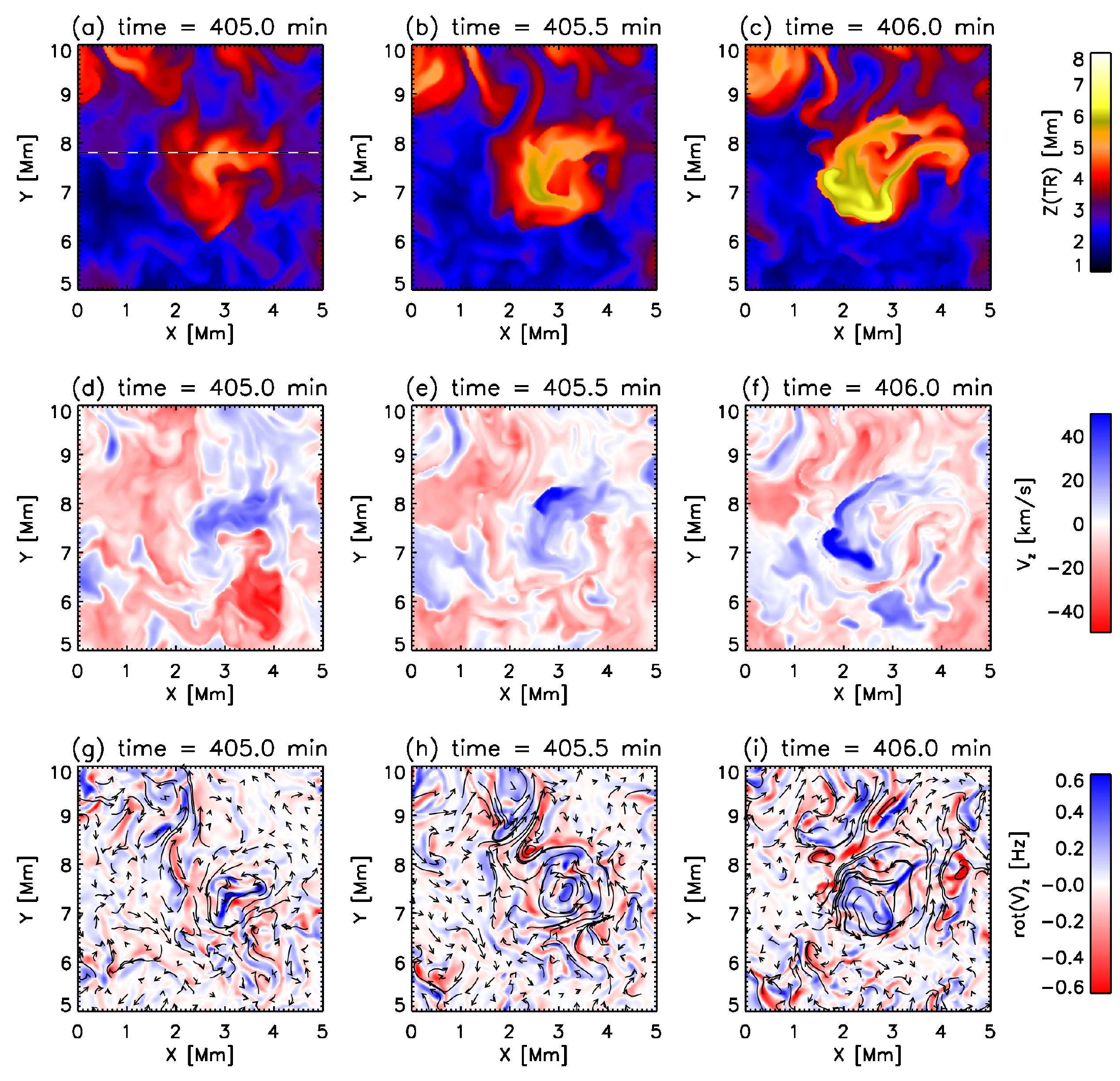}
 \caption{
 Time evolution of the transition region height (top row),
 vertical velocity (middle row),
 and vertical vorticity (middle row) at the transition region.
 Snapshots at time $=$ 405.0 (left column),
 405.5 (middle column), and 406.0 min (right column)
 are shown.
 The horizontal dashed line in the top left panel represents
 the position of the slit ($Y=7.8$ Mm) plotted in
 Figures \ref{fig:plxz_t_m}, \ref{fig:plxz_t_m2},
 and \ref{fig:plxz_force}.
 The arrows in the bottom panels indicate
 the horizontal velocity field at the transition region.
 The variables of $Y>9$ Mm are calculated by
 assuming a periodic horizontal boundary condition.
 }
 \label{fig:plxyt_ztr_zoom_m}
\end{figure}

At time $=$ 405.0 min,
Jet-A starts to rise exhibiting the anticlockwise rotation.
Figure \ref{fig:plxyt_ztr_zoom_m}
shows a closed-up view of the initial emerging phase of Jet-A.
The chromospheric jets begin to rise
in the region of $X=$ 2--4 Mm and $Y\sim$ 8 Mm
at time $=$ 405.0 min (Figures \ref{fig:plxyt_ztr_zoom_m}(a) and (d)),
showing the anticlockwise rotation
(Figure \ref{fig:plxyt_ztr_zoom_m}(g)).
This initial rising region corresponds
to the negative magnetic patch region in the chromosphere,
as shown in Figures \ref{fig:plxyt_z_m_bz.0108}(c) and (d).
The region continues to rise with the rotating motion
following the the chromospheric vortex
throughout the emergence of the chromospheric jet.
The horizontal velocity increases the complexity
during the emergence of Jet-A
(Figure \ref{fig:plxyt_z_m_bz.0108}(i)),
and produces the fine-scale structure of Jet-A.

\begin{figure}[!tp]
 \centering
 \plotone{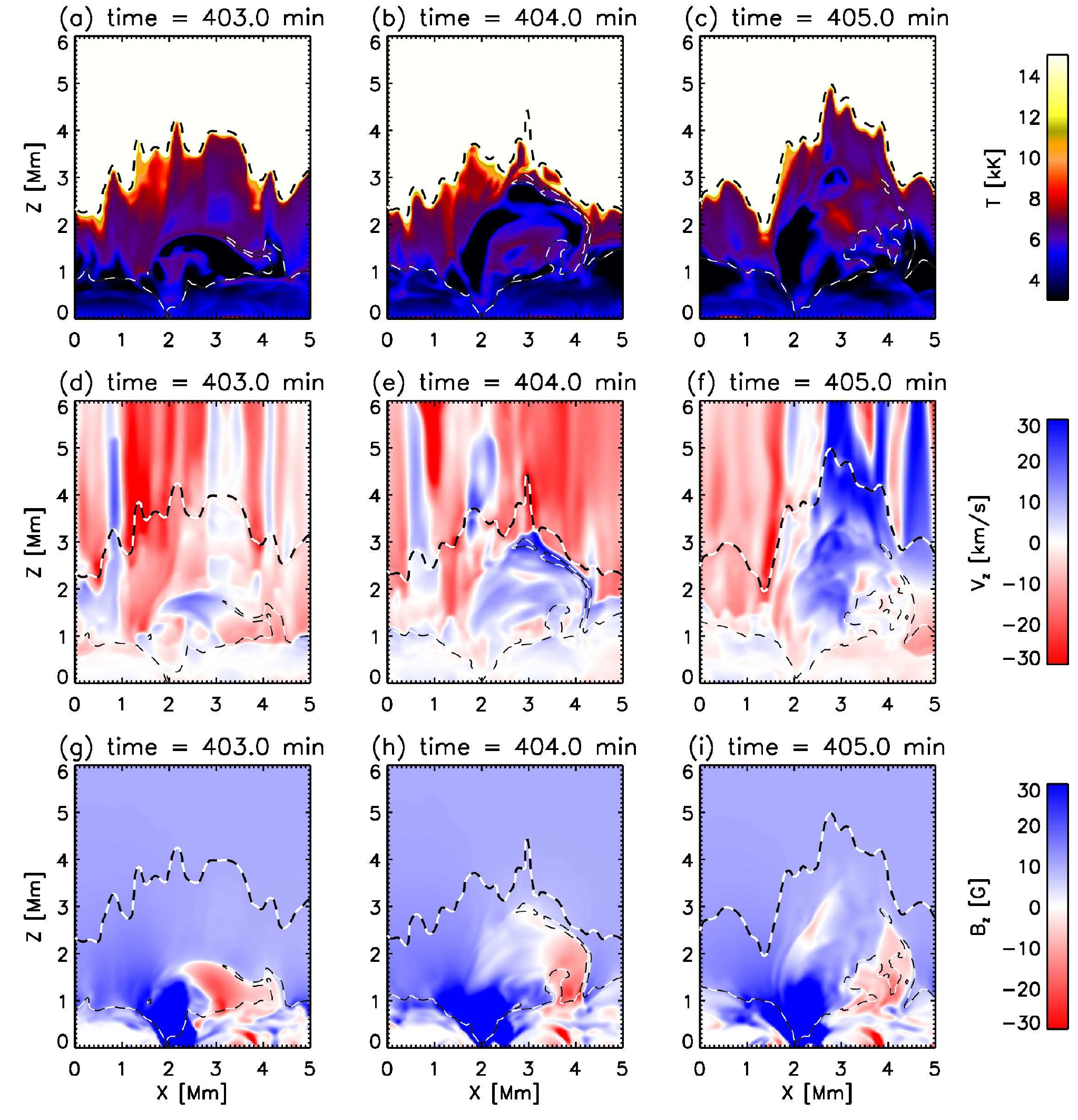}
 \caption{
 Time evolution on the $(X,Z)$-plane at $Y=7.8$ Mm.
 Shown are (a--c) the gas temperature,
 (d--f) vertical velocity,
 and (g--i) vertical magnetic field.
 The thin dashed lines indicate the plasma-beta unity.
 The thick dashed lines represent
 the position of the transition region,
 defined as the height at which the gas temperature becomes 40000 K.
 }
 \label{fig:plxz_t_m}
\end{figure}

\begin{figure}[!tp]
 \centering
 \plotone{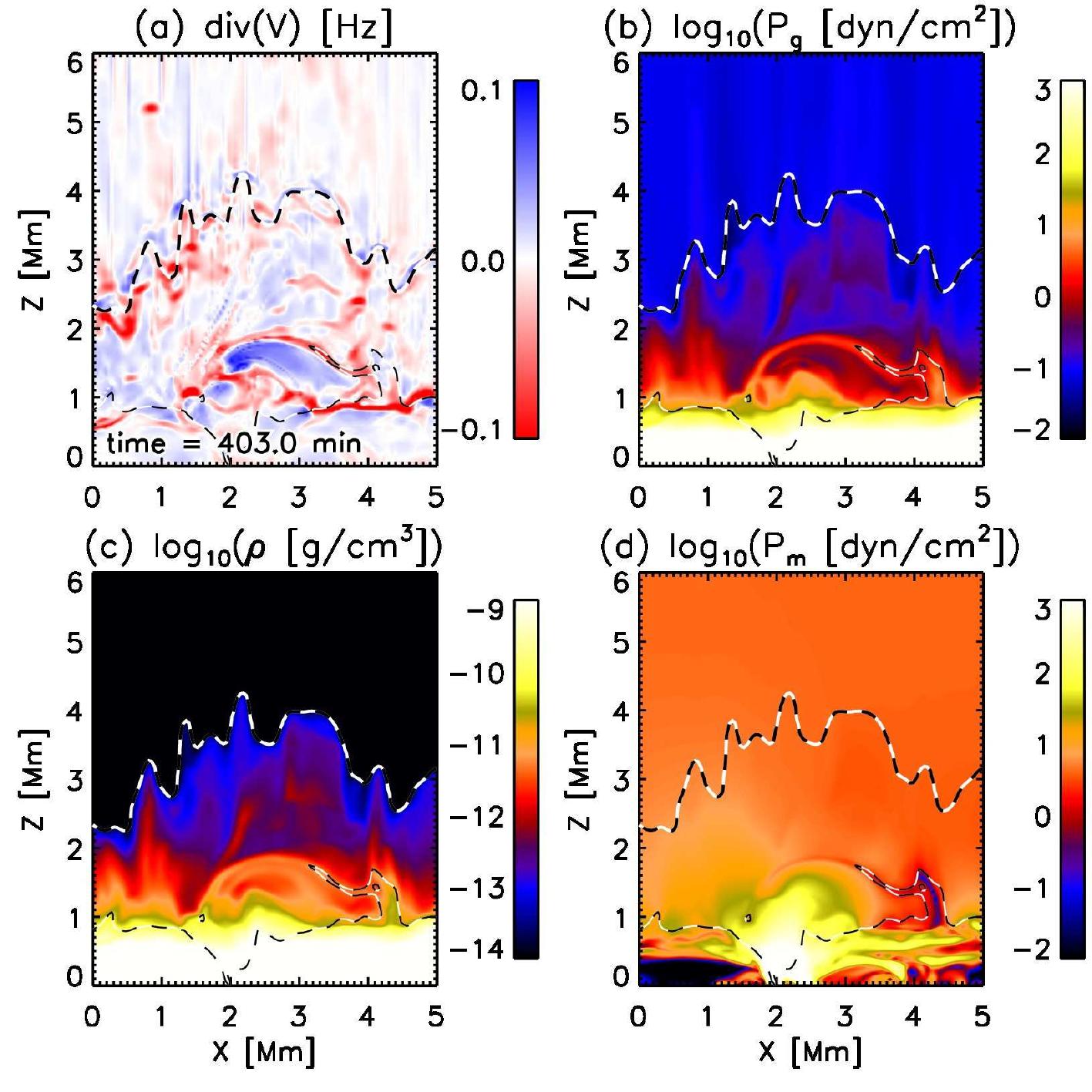}
 \caption{
 Vertical slice on the $(X,Z)$-plane of $Y=7.8$ Mm
 at time $=403.0$ min.
 Shown are (a) the divergence of the velocity field,
 (b) the gas pressure,
 (c) the mass density,
 and (d) the magnetic pressure.
 }
 \label{fig:plxz_t_m2}
\end{figure}

Before the top of Jet-A (i.e., the transition region)
starts to rise, we identify the upward motion
of a plasma blob in the chromosphere,
with the shock front at the top,
in the chromosphere.
Figures \ref{fig:plxz_t_m} and \ref{fig:plxz_t_m2}
illustrate the time evolution of the chromospheric plasma
in the driving process of Jet-A.
The vertical slice of Figure \ref{fig:plxz_t_m}
is located at the initial rising region of Jet-A
(the dashed line in Figure \ref{fig:plxyt_ztr_zoom_m}(a)).
At time $=403.0$ min,
the cool plasma blob gradually moves upward
with a vertical velocity of $\sim$ 10 km/s
(Figures \ref{fig:plxz_t_m}(a) and (d)).
The shock front is formed at the top of this plasma blob
as identified by the region with a strong velocity convergence
(Figure \ref{fig:plxz_t_m2}(a)).
Near the shock front, the gas pressure
is enhanced by compression (Figure \ref{fig:plxz_t_m2}(b)).

The region inside the plasma blob
is evacuated (i.e., low gas pressure) and highly magnetized.
Below the shock front,
the plasma blob has a lower gas pressure than the surroundings
(Figure \ref{fig:plxz_t_m2}(b)).
This reduction of the gas pressure
is caused not by the reduction of mass density
but by that of the gas temperature
(Figures \ref{fig:plxz_t_m}(a) and \ref{fig:plxz_t_m2}(c))
The magnetic field strength in this evacuated region
is higher than that in the surrounding region
(Figure \ref{fig:plxz_t_m2}(d)).
This enhancement of the magnetic pressure corresponds to
the negative magnetic patch in the chromosphere
(Figures \ref{fig:plxyt_z_m_bz.0108}(c) and (d))
caused by the twisted and deformed magnetic field
(Figures \ref{fig:pv11_0108} and \ref{fig:pv13_0108}).
This result implies that the contribution of the Lorentz force
contributes to the formation process of Jet-A.

The cool and dense plasma blob hits the transition region
and drives the tall chromospheric jet (Jet-A).
The top of the upward-moving chromospheric plasma
reaches the transition region at time $=404.0$ min
(Figures \ref{fig:plxz_t_m}(b), (e), and (h)).
The amplitude of the upward velocity exceeds 30 km/s
during the vertical propagation.
At time $=405.0$ min
(Figures \ref{fig:plxz_t_m}(c), (f), and (i)),
the shock front hits the transition region.
The upward velocity is further amplified
owing to the shock-transition region interaction
\citep{1982ApJ...257..345H}.
The chromospheric plasma continues to rise after the interaction
and forms Jet-A.
The region of the negative vertical magnetic field
located at $X\sim$ 3--4 Mm is gradually diffused
by the release of the twist of the magnetic field
during the emergence of the chromospheric jet.


\begin{figure}[!tp]
 \centering
 \plotone{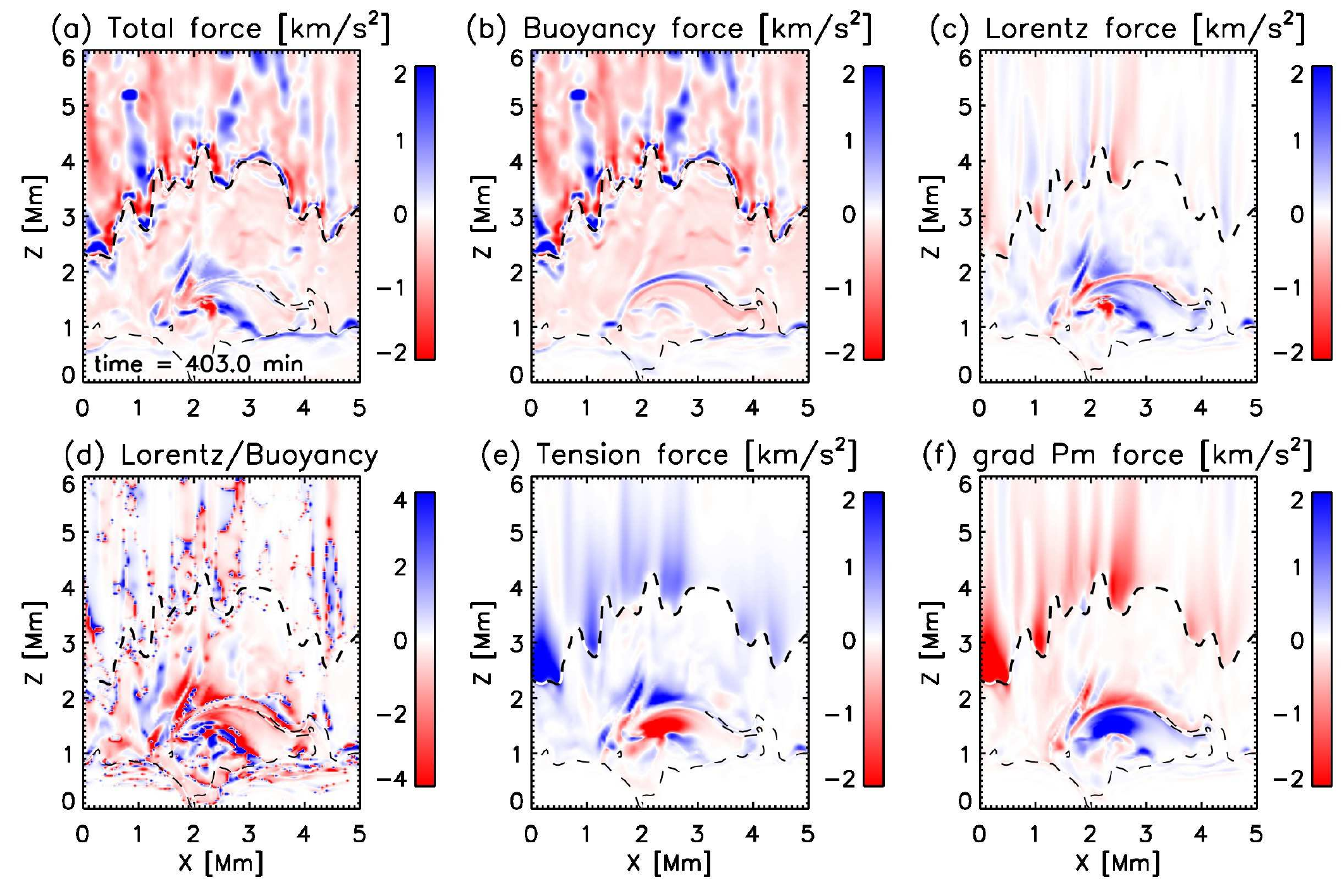}
 \caption{
 Same as Figure \ref{fig:plxz_t_m2} but showing
 the vertical components of the driving forces.
 Shown are (a) the sum of the Lorentz force
 and the buoyancy force $F_\mathrm{L}+F_\mathrm{B}$,
 (b) the buoyancy
 (gas pressure gradient plus gravitational) force $F_\mathrm{B}$,
 (c) the Lorentz force $F_\mathrm{L}=F_\mathrm{MP}+F_\mathrm{MT}$,
 (d) the ratio between the Lorentz force
 and buoyancy forces $F_\mathrm{L}/F_\mathrm{B}$,
 (e) the magnetic pressure gradient force $F_\mathrm{MP}$,
 and (f) the magnetic tension force $F_\mathrm{MT}$.
 See the body text of this paper for the definitions of
 $F_\mathrm{L}$, $F_\mathrm{B}$, $F_\mathrm{MP}$, and $F_\mathrm{MT}$.
 }
 \label{fig:plxz_force}
\end{figure}

We evaluate the vertical component of the momentum equation
to clarify the acceleration process of the cool plasma blob
(shown in Figures \ref{fig:plxz_t_m} and \ref{fig:plxz_t_m2})
which drives Jet-A.
The vertical components of
the buoyancy force $F_\mathrm{B}$
and the Lorentz force $F_\mathrm{L}$
per unit mass are given by
\begin{align}
 \label{eq:F_B}
 F_\mathrm{B}=-\frac{1}{\rho}\frac{\partial P}{\partial z}+g_z
\end{align}
and
\begin{align}
 \label{eq:F_L}
 F_\mathrm{L}=\frac{1}{4\pi\rho}\left[
 \left(\nabla\times\bm{B}\right)\times\bm{B}
 \right]_z,
\end{align}
respectively.
We decompose the Lorentz force $F_\mathrm{L}$
into the magnetic pressure gradient force $F_\mathrm{MP}$
and the magnetic tension force $F_\mathrm{MT}$
for a clearer interpretation.
These forces are written as
\begin{align}
 \label{eq:F_MP}
 F_\mathrm{MP}=-\frac{1}{\rho}\frac{\partial P_m}{\partial z}
 +\frac{b_z}{\rho}\left(\bm{b}\cdot\nabla\right)P_m
\end{align}
and
\begin{align}
 \label{eq:F_MT}
 F_\mathrm{MT}=F_\mathrm{L}-F_\mathrm{MP},
\end{align}
respectively, where $P_m=B^2/(8\pi)$ is the magnetic pressure
and $\bm{b}=\bm{B}/B$ is the unit vector
in the direction of the magnetic field.

We find that the upward motion of the cool plasma blob
and the resulting chromospheric jet (Jet-A)
is driven by the Lorentz force,
especially by the magnetic pressure gradient force.
Figure \ref{fig:plxz_force} shows
the vertical slice of the driving forces.
The buoyancy force mainly acts downward
except near the shock front
where the pressure gradient force acts
both upward and downward (Figure \ref{fig:plxz_force}(b)).
The Lorentz force is relatively strong
and mainly acts upward (\ref{fig:plxz_force}(c)).
As shown in Figure \ref{fig:plxz_force}(d),
the Lorentz force is the dominant driver
of the upward motion of the cool chromospheric plasma blob
(Figures \ref{fig:plxz_t_m} and \ref{fig:plxz_t_m2}).
In particular,
the magnetic pressure gradient force $F_\mathrm{MP}$
acts upward at the plasma blob (Figure \ref{fig:plxz_force}(e)).
The magnetic tension force (Figure \ref{fig:plxz_force}(f))
tends to compensate for the magnetic pressure gradient force.
In total (Figure \ref{fig:plxz_force}(a)),
the left part of the cool plasma blob
is accelerated upward by the magnetic pressure gradient force.
From the above results,
we conclude that the chromospheric shock wave,
as an energy source of Jet-A,
is driven by the magnetic pressure gradient force.

We will now summarize the formation process of
the tallest chromospheric jet in our simulation (Jet-A):
\begin{enumerate}
 \item First, the twisted magnetic field structure is generated
       by the vortex motion at the root of the jet
       (Figure \ref{fig:pv11_0108}).
       A part of the magnetic field lines is dragged and deformed
       by the lower chromospheric plasma
       owing to the inclination of the vortex away
       from the vertical axis (Figure \ref{fig:plxyt_z_m_vh}).
 \item After a sufficiently large magnetic twist is stored,
       these deformed magnetic field lines
       start to lift up the dense and cool chromospheric plasma
       to the upper layer by the Lorentz force
       (Figures \ref{fig:pv13_0108} and \ref{fig:plxz_force}).
 \item The accelerated chromospheric plasma
       exceeds the sound speed and becomes a shock wave
       (Figure \ref{fig:plxz_t_m2}).
       The shock wave hits the transition region
       and finally forms the chromospheric jet
       (Figure \ref{fig:plxz_t_m}).
\end{enumerate}

\section{Discussion}\label{sec:discuss}

We briefly compare our simulation with the observed chromospheric jets.
In this paper,
we present the results of a three-dimensional simulation
of solar chromospheric jets.
We successfully reproduce a chromospheric jet (Jet-A)
with a maximum height of 10--11 Mm
and lifetime of 8--10 min.
The tall chromospheric jet in our simulation
shows a parabolic path with 
a maximum velocity of approximately 50 km/s,
and a deceleration of approximately 200 m/s$^2$.
These parameters are in good agreement with
the observed properties of the spicules
(or more precisely the classical Type I spicules) in quiet regions
\citep{2015ApJ...806..170S}.
These jets emanate from a strong magnetic flux tube below
and are driven by a Lorentz force
related to the rotational motion of the flux tube.


The driving mechanism and resulting jet in our simulation
are similar to those in the one-dimensional Alfv\'en wave models
of chromospheric jets presented in the previous studies,
but they also exhibit several differences.
The chromospheric jet described in this paper
originates from the vortex motion in the lower atmosphere.
In this sense, our model is categorized
as a family of the Alfv\'en wave models of spicules
\citep{1982SoPh...75...35H,1999ApJ...514..493K}.
However, there are several differences from previous studies
concerning the driving mechanism and resulting jets
caused by the axial asymmetry of the vortex
in the three-dimensional domain.

One such differences from the one-dimensional models
is that the inclination of the vortex tube
also plays a role in the driving process of our simulation.
We show that the inclined rotation of the magnetic field lines
lifts the dense chromospheric plasma upward.
In one-dimensional the Alfv\'en wave models,
the nonlinearity or amplitude of the Alfv\'en wave
is important for the generation of chromospheric jets,
since the highly nonlinear Alfv\'en wave
is required to produce the acoustic wave (or shock wave)
through the mode conversion \citep{1971JGR....76.5155H}.
We suggest that the inclination of the magnetic flux tube
may also helps in increasing the twist of the magnetic field
(and the nonlinearity of the Alfv\'en wave) in the chromosphere.



Another large difference between our study
and one-dimensional axisymmetric Alfv\'en wave models
is that the chromospheric jet
has a fine-scale axially asymmetric structure.
The produced chromospheric jet
forms a cluster with a diameter of several Mm
that consists of the finer strands
as shown in Figures \ref{fig:plztr_m} and \ref{fig:plamity_m}.
This result is consistent with the multi-threaded nature
of spicules \citep{2008ASPC..397...27S,
2010ApJ...714L...1S,2014ApJ...795L..23S}.
We also find that the horizontal size of the cluster of strands
gradually becomes wider during the evolution of the jet.
This is probably related to
the centrifugal force of the rotation of the jet.
The rotation of the whole cluster
with the independent separating motion of finer strands
found in our simulation
reminds us of the separation and connection of the spicule strands
and the spatial broadening and diffusion of spicules
\citep{2010ApJ...714L...1S}.
\cite{2014ApJ...795L..23S}
suggested that the Kelvin-Helmholtz instability
caused by the swaying motion
produces this multi-threaded nature
as in the coronal loop simulations \citep{2014ApJ...787L..22A}.
The difference is that our chromospheric jets
are driven by rotational rather than swaying motion.
\cite{2008ASPC..397...27S} interpreted the separation of threads
by assuming the torsional motion of a spicule as a rigid body.
Although their interpretation also works in our model,
our horizontal structure seems to be
closely related to the jet's formation process.
It is not easy to determine the main contributor
of the fine-scale horizontal structure formation
from the several candidates such as
the horizontal inhomogeneity of the initial driver
or latter Kelvin-Helmholtz instability.
We also note that our pseudo-emission,
shown in Figure \ref{fig:plamity_m},
is not the real radiative emission observed in the solar chromosphere.
Under the limitations described above,
our current interpretation of the formation process
of the horizontal structure in Jet-A is as follows:
\begin{enumerate}
 \item Initially, the large-scale (anticlockwise) rotation
       with a typical diameter of 0.5--1.0 Mm is
       driven by the release of the magnetic twist in the chromosphere
       (Figure \ref{fig:plxyt_z_m_vh}).
 \item During the evolution, the strong velocity shear
       between Jet-A and the surrounding coronal plasma
       causes the Kelvin-Helmholtz instability
       and forms a multi-strand structure
       with smaller vortices with a typical diameter smaller
       than several hundreds of kilometers.
       The small-scale clockwise rotation
       (Figures \ref{fig:plxyt_ztr_zoom_m})
       is generated in this stage.
       The initially small phase difference
       of the Alfv\'en wave will be also enhanced,
       helping in forming the fine-scale structure.
 \item Resulting fine-scale strands are
       separated by the centrifugal force
       of the initial large-scale anticlockwise rotation
       (Figures \ref{fig:plztr_m} and \ref{fig:plamity_m}).
\end{enumerate}
Further analysis is required to clarify the exact mechanism
of horizontal structure formation and
the relationship between the radiative emission
and the horizontal structure of the chromospheric plasma.


We have shown that the simulated chromospheric jet
exhibits the apparent horizontal oscillation
during its lifetime (Figure \ref{fig:plamity_xt}).
The period of several minutes
and the amplitude of 20-30 km/s
are consistent with the observational
oscillatory nature of solar spicules
\citep{2009SSRv..149..355Z}.
In our model, the initial anticlockwise torsional motion
and the small-scale vortices of the fine-scale strands
are the origin of this apparent oscillation.
It should be noted that the apparent periodicity
of (pseudo-)radiative emission does not indicate
the periodic motion of the actual plasma motion.
The combination of the fine-scale structure inside the jet
and the vortex is the origin of the apparent periodicity or oscillation.
One important aspect of our model is that the density
and horizontal velocity structure are not axisymmetric.
This reveals the importance of considering
the non-axisymmetric model
to understand the oscillation of observed spicules.
Since we use a very rough approximation
to compute the (pseudo-)radiative emission,
a more detailed analysis
should be carried out to
compare Jet-A with the observation.
The propagation of the produced oscillation
should also be investigated in the future.


We find a strong vortex at the root of the tall chromospheric jet.
The tornado-like streamlines above the strong magnetic concentration
have been reported by \cite{2012Natur.486..505W}
and numerically investigated by several authors
\citep{2012Natur.486..505W,2013ApJ...770...37K,
2013ApJ...776L...4S,2014PASJ...66S..10W,2017A&A...601A.135K}
without including the corona above the chromosphere.
In our simulation,
the chromospheric  magnetic field is also twisted
following the swirling motion.
\cite{2013ApJ...770...37K} reported weakly twisted
magnetic field lines produced by chromospheric swirls
with an average vertical magnetic field strength of 10 G,
which is exactly the same as that in our case.
A hot center with a downward flow
and a cool edge with an upward flow
of their swirl event
are also found in both their swirl and our simulation.
The existence of chromospheric jets
and the highly twisted magnetic field lines
is probably produced
by the inclusion of the corona above the chromosphere.
The simulations conducted by \cite{2012Natur.486..505W}
and \cite{2013ApJ...776L...4S}
investigated the cases with a stronger magnetic field.
Both of studies exhibited little twist of the magnetic field
probably owing to the high Alfv\'en speed in the chromosphere.
The effect of the average vertical magnetic field strength
on the formation of chromospheric jets
should be investigated in the future.


One of the limitations of our model is
the spatial grid size of the simulation.
We find that, in the two-dimensional simulation
\citep{2015ApJ...812L..30I},
the width or the horizontal size of the jets
becomes finer when a smaller grid size is used.
Similar phenomena can occur in our model.
The rotational motion driven in the photosphere
and the upper convection zone
also require spatial resolution
because the magnetic field is intensified
into small flux tubes in the photosphere.
Note that our simulation does not produce
tall jets such as reported in this study
if we double the horizontal grid size
(low spatial resolution) used for the simulation.

Another limitation of our model is the approximations
in the equations of state and the radiative cooling.
The LTE equation of states
and simplified radiative cooling are used in this study.
The large amplitude of radiative cooling
reduces the length and maximum velocity
of the produced chromospheric jets
by damping the chromospheric acoustic/shock waves
\citep{1990ApJ...349..647S,2013ApJ...766..128G}.
The assumption of the LTE equation of states in the chromosphere
causes an error in the heat capacity (or gas temperature)
and the electron number density.
The effect of non-equilibrium ionization
\citep{2002ApJ...572..626C,2006A&A...460..301L,
2007A&A...473..625L,2011A&A...528A...1W,
2014ApJ...784...30G,2016ApJ...817..125G}
on the formation of chromospheric jets
should be investigated in the future.

The effect of the collisions between the ions and neutrals
is another issue to be studied.
This effect appears in the Ohm's law as additional terms
in the one-fluid MHD equations.
In our calculations here, no explicit resistivity
besides the numerical diffusivity is introduced.
The importance of the Hall effect or the ambipolar diffusion
in the solar chromosphere is suggested by the previous studies
\citep[e.g.,][]{2012ApJ...753..161M,2012ApJ...750....6C,
2014SSRv..184..107L,2015RSPTA.37340268M,2017PPCF...59a4038K}
More recently, \cite{Martinez-Sykora1269} have suggested
that the ambipolar diffusion helps in increasing
the length of chromospheric jets driven by the tension force.
As we have shown,
the dragging of the chromospheric plasma
by the torsional motion plays an important role
in the production of chromospheric jets.
In the real solar chromosphere,
the ambipolar diffusion produced by the drift motion
between the ions and neutrals
will reduce the dragging efficiency of the chromospheric plasma,
leading to shorter chromospheric jets.
A similar reduction in the chromospheric plasma drag
has been investigated in the context of flux emergence
and active region formation
\citep{2006A&A...450..805L,2007ApJ...666..541A,2013ApJ...764...54L}.
Another effect of the ambipolar diffusion
is its assistance in the formation of thin current sheets
\citep{1963ApJS....8..177P,1994ApJ...427L..91B}
and the occurrence of fast magnetic reconnection
in the chromosphere.
It should also be noted that,
because the amount of the ambipolar diffusion strongly
depends on the ionization rate,
the detailed treatment of the equation of state
\citep{2007A&A...473..625L,2016ApJ...817..125G}
or the amount of the radiative cooling
\citep{laguna2017effect}
will cause significant differences on the simulation results.
Further work with a greater
focus on the ambipolar diffusion should be undertaken.

We find a chromospheric jet that is significantly taller than
those in similar three-dimensional radiative MHD simulations
\citep{2009ApJ...701.1569M,2011ApJ...736....9M,2015ApJ...811..106H}.
These simulations employed the Oslo Stagger code
or the Bifrost code \citep{2011A&A...531A.154G}
with a more realistic treatment of the radiation.
Because of the many differences between our code (RAMENS) and Bifrost,
it is not easy to understand what causes this difference.
One candidate is the assumption of closed loop configuration
for the magnetic field in their study.
This possibly acts to reduce the maximum height
of the chromospheric jets that extend along the loop.
The closed loop also acts to heat the corona more easily.
This prevents the nonlinear amplification
of the chromospheric shock wave
\citep{1982SoPh...78..333S,2015ApJ...812L..30I},
leading to shorter jets.
The numerical dissipation of the MHD scheme is also important.
As discussed above, we do not find tall chromospheric jets,
like Jet-A, at a lower spatial resolution.
\cite{2015ApJ...811..106H} obtained results with different grid sizes
and showed that the chromospheric material is more elongated
for a finer spatial grid.
The treatment of the radiative cooling is
also a possible cause of the difference.
Stronger radiative cooling causes stronger damping
of the chromospheric shock waves.
Therefore, the resulting chromospheric jets will be shorter.
A more detailed comparison is required to show
the dominant source of difference
between our results and those of previous studies.

\section{Conclusion}

We have presented a radiation magnetohydrodynamic simulation
of chromospheric jets.
A tall chromospheric jet (Jet-A)
is reproduced by the Lorentz force
of the twisted magnetic field lines.
Jet-A also exhibits horizontal oscillatory motion
and fine-scale internal structure during its emergence.
Our model is a three-dimensional extension
of the classical Alfv\'en wave model of chromospheric jets.
In this study, the most significant characteristic of the model
is that the all of the driving mechanism,
multi-strand structure, and oscillatory nature
are closely related to each other and explained in a model.
We conclude that the Alfv\'en wave model,
or the torsional wave model of spicules,
is an important candidate for explaining
the driving mechanism of solar spicules.

\acknowledgments

This work was supported by JSPS KAKENHI Grant Number JP15H05816, JP15H03640,
``Joint Usage/Research Center for
Interdisciplinary Large-scale Information Infrastructures'',
``High Performance Computing Infrastructure'',
and the Program for Leading Graduate Schools, MEXT, in Japan.
Numerical computations were carried out
on the Cray XC30 supercomputer at the Center
for Computational Astrophysics, National Astronomical Observatory of Japan.
The authors are grateful to the anonymous referee
for improving the manuscript.






\end{document}